\documentclass[preprint,superscriptaddress,nofootinbib]{revtex4}
     
     \usepackage{amssymb,amsmath,graphicx,color}
     \usepackage[colorlinks]{hyperref}
     \usepackage{multirow,natbib,rotating}
     \begin{document}
     \title{Study of the two-body nonleptonic $B_{c}(2S)$ weak decays
      with the QCD factorization approach}
     \author{Na Wang}
     \email[corresponding author, ]{wangna@haut.edu.cn}
     \affiliation{School of Physics and Advanced Energy,
                 Henan University of Technology, Zhengzhou 450001, China}
     \author{Yueling Yang}
     \email[corresponding author, ]{yangyueling@htu.edu.cn}
     \affiliation{School of Physics,
                 Henan Normal University, Xinxiang 453007, China}
     \author{Junfeng Sun}
     \email[corresponding author, ]{sunjunfeng@htu.edu.cn}
     \affiliation{School of Physics,
                 Henan Normal University, Xinxiang 453007, China}

     \begin{abstract}
     Inspired by the promising prospect of the
     $B_{c}(2S)$ meson at the coming HL-LHC experiments,
     the nonleptonic $B_{c}(2S)$ meson weak decays induced by
     both the $b$ and $c$ decays are investigated with the QCD
     factorization approach.
     It is found that branching ratios for the color- and CKM-favored
     $B_{c}(2S)$ ${\to}$ $B_{s}{\rho}$, $B_{s}{\pi}$ decays can
     reach up to ${\cal O}(10^{-9})$, which might be measurable.
     The $B_{c}(2S)$ decays into the final states containing one
     charmonium are highly suppressed by the CKM factors, and have
     significantly small branching ratios, ${\cal O}(10^{-12})$
     and less, which might be outside the future measurement capability.
     This paper provides a ready reference for the future
     experimental study on the hadronic $B_{c}(2S)$
     weak decays.
     \end{abstract}
     \maketitle

     \section{Introduction}
     \label{sec01}
     The $B_{c}$ meson family is very unique within the quark
     model assignments.
     The $B_{c}$ mesons are composed of two heavy quarks $b$ and $c$
     with different flavors, and have nonzero bottomness ($B$)
     and charm ($C$) and electric charge ($Q$) quantum numbers,
     $B$ $=$ $C$ $=$ $Q$ $=$ ${\pm}1$, no matter what configuration
     of the spin and relative orbital angular momentum of
     two quarks.
     The charged ``self-tagging'' $B_{c}$ mesons should in
     principle be easily identified at experiments.
     However, up to now, only two established pseudoscalar $B_{c}$
     mesons with the same isospin and spin and parity quantum numbers
     $I(J^{P})$ $=$ $0(0^{-})$ among the abundant spectrum have
     been clearly listed in the latest PDG \cite{PhysRevD.110.030001},
     {\em i.e.}, the ground $S$-wave spin-singlet $B_{c}(1S)$
     meson and its first radial excited state marked as
     the $B_{c}(2S)$ meson.

     Theoretically, besides the direct production through virtual
     ${\gamma}^{\ast}$ and $Z^{\ast}$ boson exchanges in $e^{+}e^{-}$
     annihilation and the indirect production in $Z^{0}$, $W^{\pm}$,
     $H^{0}$ boson decays and $t$ quark decays
     \cite{PhysLettB.284.127,PhysRevD.46.3845,EPJC.71.1563,
     PhysRevD.85.094015,EPJC.82.246,PhysRevD.106.094036,
     PhysLettB.348.219,ZPC.63.77,PhysLettB.342.351,PhysLettB.381.341,
     NPB.915.224,NPB.911.36,JHEP.2011.08.131,PhysRevD.91.034033,
     PhysRevD.102.016011,CPC.42.083107,EPJC.71.1766,
     PhysRevD.101.036009,PhysRevD.101.034029,PhysRevD.93.054031,
     PhysRevD.98.036014,PhysRevD.100.053002,NPA.1015.122285,
     PhysRevD.107.074005,PhysRevD.54.5606,PhysRevD.81.114035,
     PhysRevD.77.014022,EPJC.78.657},
     the $B_{c}$ mesons can be produced principally in hadron-hadron collisions
     by both the gluon–gluon fusion mechanism and quark–antiquark
     annihilation \cite{PhysRevD.48.4086,0412158,EPJC.38.267}.
     The gluon–gluon fusion mechanism is dominant over quark–antiquark
     annihilation at the Tevatron and 
     LHC \cite{0412158}.
     Both the bottomness and charm quantum numbers should be
     strictly and individually conserved in the chromatic interactions.
     A $b\bar{b}$ quark pair plus a $c\bar{c}$ quark pair must be
     generated simultaneously, in which a $b$ (or $\bar{b}$) quark
     and a $\bar{c}$ ($c$) quark should be extremely fortunate to
     bind together and finally hadronize into a color singlet
     $B_{c}$ bound state.
     At the leading order in ${\alpha}_{s}$, the production cross section
     in hadron-hadron collisions is proportional to ${\alpha}_{s}^{4}$
     and so significantly small and strongly dependent on the collision
     energy \cite{PhysRevD.48.4086,0412158,EPJC.38.267,PhysRevD.97.114022},
     where ${\alpha}_{s}$ is the coupling of the strong interactions.
     It was estimated that the inclusive production cross section of
     the $B_{c}(1S)$ meson at the LHC is at a level of $1\,{\mu}{\rm b}^{-1}$
     for $\sqrt{s}$ ${\approx}$ $14$ TeV, about one order of magnitude larger
     than that at the Tevatron \cite{0412158,EPJC.38.267,PhysRevD.97.114022,CPL.27.061302}.
     This means that ${\cal O}(10^{9})$ $B_{c}(1S)$ mesons can
     be anticipated with $1\, {\rm fb}^{-1}$ of accumulated
     data at the LHC, and more than $10^{12}$ $B_{c}(1S)$ mesons
     will be available with a prospective integrated luminosity
     of $4\, {\rm ab}^{-1}$ at the HL-LHC 
     during its exploitation period \cite{web.HL-LHC}.
     Correspondingly, there will be approximately ${\cal O}(10^{11})$
     $B_{c}(2S)$ available at HL-LHC experiments,
     using the ratio of the $B_{c}(2S)$ to $B_{c}(1S)$ production cross
     sections at $\sqrt{s}$ $=$ $13$ TeV measured by the CMS group,
     $R$ $=$ $\displaystyle \frac{ {\sigma}(B_{c}(2S)) }{ {\sigma}(B_{c}(1S)) }\,
     {\cal B}r(B_{c}(2S){\to}B_{c}(1S){\pi}^{+}{\pi}^{-})$ $=$
     $(3.47{\pm}0.63{\pm}0.33)\%$ \cite{PhysRevD.102.092007}.
     A vast amount of promising data provides a solid experimental
     foundation and precious opportunity to investigate thoroughly
     the properties of the $B_{c}(1S)$ and $B_{c}(2S)$ mesons.

     Experimentally, compared with the $B_{c}(1S)$ meson,
     the $B_{c}(2S)$ meson is still an immature and up-and-coming
     particle, who is first discovered in the
     $B_{c}(1S){\pi}^{+}{\pi}^{-}$ invariant-mass spectrum
     by the ATLAS group in 2014 \cite{PhysRevLett.113.212004}
     and then identified by both the CMS \cite{PhysRevLett.122.132001}
     and LHCb \cite{PhysRevLett.122.232001} groups in 2019.
     The properties of the $B_{c}(1S)$ meson has been extensively
     studied whether at experiment or theory.
     In contrast, our knowledge about the properties of the
     $B_{c}(2S)$ meson is woefully inadequate and urgently needs
     to be enhanced and enriched further.
     In the dictionary of the authoritative PDG \cite{PhysRevD.110.030001},
     the averaged mass of the $B_{c}(2S)$ meson is $m$ $=$ $6871.2$
     ${\pm}$ $1.0$ MeV and lies below the open $BD$ threshold.
     The favourite decays, predominantly by hadronic cascades and the
     electric dipole (E1) or magnetic dipole (M1) transitions,
     lead to the total width of the $B_{c}(2S)$ meson are generally
     less than one hundred keV \cite{PhysRevD.49.5845,PhysRevD.60.074006,
     PhysRevD.70.054017,PhysRevD.100.096002,PhysRevD.106.054009,EPJC.83.1080},
     some of which are listed in Table \ref{width-bc2s}.
     It is easy to imagine that the reconstruction of demonstrable
     signals of the hadronic decay $B_{c}(2S)$ ${\to}$
     $B_{c}(1S){\pi}^{+}{\pi}^{-}$ at experiments should be
     very challenging, because $m_{B_{c}(2S)}$ $-$ $m_{B_{c}(1S)}$
     $-$ $m_{2\,{\pi}}$ ${\approx}$ $320$ MeV and the soft pions
     are submerged in noisy background.
     For the $B_{c}(2S)$ electromagnetic transitions, both the final
     $B_{c}(1P^{(\prime)})$ and $B_{c}^{\ast}(1S)$ states have not
     been identified definitely at experiments for the moment.
     Besides, the $B_{c}(2S)$ meson can still decay via the weak
     interactions.

     \begin{table}[h]
     \caption{The partial widths (${\Gamma}_{i}$ in unit of keV) and
              branching ratios (${\cal B}r_{i}$ in unit of \%)
              of the $B_{c}(2S)$ meson decay, where the excited
              axion-like $1P$ and $1P^{\prime}$ particles are the
              weighted mixtures of the $P$-wave spin triplet ${}^3P_{1}$
              and singlet ${}^1P_{1}$ states with the same spin-parity
              $J^{P}$ $=$ $1^{+}$ quantum numbers;
              the vector $B_{c}^{\ast}(1S)$ meson is the ground
              $S$-wave spin triplet ${}^3S_{1}$ state with the
              spin-parity $J^{P}$ $=$ $1^{-}$ quantum numbers;
              the total width of the $B_{c}(2S)$ meson is
              approximately the sum of four partial widths
              listed here.}
     \label{width-bc2s}
     \begin{ruledtabular}
     \begin{tabular}{cc cc cc cc cc cc cc}
       decay & final
     &  \multicolumn{2}{c}{ Ref. \cite{PhysRevD.49.5845} }
     &  \multicolumn{2}{c}{ Ref. \cite{PhysRevD.60.074006} }
     &  \multicolumn{2}{c}{ Ref. \cite{PhysRevD.70.054017} }
     &  \multicolumn{2}{c}{ Ref. \cite{PhysRevD.100.096002} }
     &  \multicolumn{2}{c}{ Ref. \cite{PhysRevD.106.054009} }
     &  \multicolumn{2}{c}{ Ref. \cite{EPJC.83.1080} }
     \\ \cline{3-4} \cline{5-6} \cline{7-8}
        \cline{9-10} \cline{11-12} \cline{13-14}
       mode  & states
     & ${\Gamma}_{i}$ & ${\cal B}r_{i}$
     & ${\Gamma}_{i}$ & ${\cal B}r_{i}$
     & ${\Gamma}_{i}$ & ${\cal B}r_{i}$
     & ${\Gamma}_{i}$ & ${\cal B}r_{i}$
     & ${\Gamma}_{i}$ & ${\cal B}r_{i}$
     & ${\Gamma}_{i}$ & ${\cal B}r_{i}$  \\ \hline
     hadronic & $B_{c}(1S) \, {\pi} \, {\pi}$     & $50~$ & $90.4$   & $50~$  & $71.9$   & $57~$ & $88.1$   & $10.68$ & $51.42$   & $42~$ & $54.4$   & $25~$ & $76.9$  \\
     E1       & $B_{c}(1P) \, {\gamma}$           & $0.0$ & $~0.0$   & $~6.4$ & $~9.2$   & $1.3$ & $~2.0$   & $~7.11$ & $34.23$   & $0.0$ & $~0.0$   & $2.8$ & $~8.6$  \\
     E1       & $B_{c}(1P^{\prime}) \, {\gamma}$  & $5.2$ & $~9.4$   & $13.1$ & $18.8$   & $6.1$ & $~9.4$   & $~2.91$ & $14.01$   & $35~$ & $45.3$   & $4.4$ & $13.5$  \\
     M1       & $B_{c}^{\ast}(1S) \, {\gamma}$    & $0.1$ & $~0.2$   & $0.06$ & $~0.1$   & $0.3$ & $~0.5$   & $~0.07$ & $~0.34$   & $0.25$ & $~0.3$  & $0.3$ & $~1.0$  \\ \hline
     total    &                                   & $55.3$ & $100$   & $69.6$ & $100$    & $64.7$ & $100$   & $20.77$ & $100$     & $77.3$ & $100$   & $33~$ & $100$
     \end{tabular}
     \end{ruledtabular}
     \end{table}

     The $B_{c}(2S)$ meson weak decays have not attracted necessary
     attention, and so there was very few research work on the
     $B_{c}(2S)$ weak decays.
     Theoretically, the weak decays of the $B_{c}(2S)$ meson are
     similar to those of the lowest $B_{c}(1S)$ meson that can
     decay only through the weak interactions within the standard
     model (SM) of elementary particles, and can be
     categorized into three classes \cite{ZPC.51.549,PhysRevD.49.3399,
     PLB.371.105,PhysRevD.77.074013,PhysRevD.77.114004,
     PhysRevD.81.074012,PhysRevD.89.114019,Sci.CN.57.1891,
     AHEP.2015.104378,CN.Sci.Bull.59.3748}:
     (1) the $b$ quark decay with the $c$ quark as a spectator,
     (2) the $c$ quark decay with the $b$ quark as a spectator, and
     (3) the $b$ and $c$ quarks annihilation into a virtual
         $W^{\pm}$ boson.
     It is estimated that the branching ratio of the $B_{c}(2S)$
     meson weak decays is tiny, about
     ${1}/ {\Gamma}_{B_{c}(2S)} \, {\tau}_{B_{c}(1S)}$
     ${\sim}$ ${\cal O}(10^{-8})$,
     where ${\tau}_{B_{c}(1S)}$ and ${\Gamma}_{B_{c}(2S)}$ are
     the lifetime of the $B_{c}(1S)$ meson and the total
     width of the $B_{c}(2S)$ meson, respectively.
     An observation of an abnormally large production rate of
     the $B_{c}(2S)$ meson weak decays would be a hint of new
     physics beyond SM.
     An advantage of weak decays compared with the strong and
     electromagnetic transitions is that
     the $B_{c}(2S)$ meson might be easily identified at
     experiments from its secondary particles having relatively
     larger momentum at the rest $B_{c}(2S)$ frame.
     Furthermore, there are much richer weak decay modes,
     which could be used to over-constrain the parameters
     obtained from the $B$ and $D$ meson decays,
     and check the applicability of various phenomenological
     models.
     Based on the potential prospects of the $B_{c}(2S)$ meson
     in the planed LH-LHC, it seems that there are some
     possibilities for experimental study on the $B_{c}(2S)$
     meson weak decays.

     In this paper, we will estimate the branching ratios for the
     nonleptonic two-body $B_{c}(2S)$ weak decays using the
     phenomenological QCD factorization (QCDF) approach
     \cite{PhysRevLett.83.1914,
     NuclPhysB.591.313,NuclPhysB.606.245,PhysLettB.488.46,
     PhysLettB.509.263,PhysRevD.64.014036},
     focusing on the $B_{c}(2S)$ ${\to}$ ${\psi}(nS)\,P$,
     ${\eta}_{c}(nS)\,P$ decays induced by the $b$ ${\to}$ $c$
     quark transition belonging to the class (1),
     and the $B_{c}(2S)$ ${\to}$ $B\,P$, $B\,V$ decays induced
     by the $c$ quark decays belonging to the class (2),
     where $P$ and $V$ denote the light ground pseudoscalar
     and vector mesons.
     The $B_{c}(2S)$ weak annihilation decays belonging to the
     class (3) are power suppressed with the QCDF approach.
     It is estimated that the dominant contribution to the lifetime
     of the $B_{c}(1S)$ meson is given by the $c$ quark decays
     (${\approx}$ $70\,\%$) and the $b$ quark decays
     (${\approx}$ $20\,\%$) \cite{0412158}.
     The class (1) and (2) decays can be easily distinguishable
     on the diagrammatic level, and they do not interfere with each
     other in any appreciable way \cite{PLB.371.105}.
     Our study will provide a ready and helpful reference
     for experimental discovery and investigation.
     The remaining parts of this paper are as follows.
     The Section \ref{sec02} delineates the theoretical framework
     for the nonleptonic $B_{c}(2S)$ weak decays.
     The Section \ref{sec03} is dedicated to presenting the numerical
     results and discussions.
     The Section \ref{sec04} is a brief summary.

     \section{theoretical framework}
     \label{sec02}

     \subsection{The effective Hamiltonian}
     \label{sec0201}
     For the $B_{c}(2S)$ ${\to}$ $B\,P$, $B\,V$ decays induced
     by the $c$ quark decays, the effective weak interaction
     Hamiltonian can be written as \cite{PhysRevD.77.114004},
     \begin{equation}
    {\cal H}_{\rm eff}^{c} \, = \,
     \frac{G_{F}}{\sqrt{2}}
     \sum\limits_{q_{1},q_{2}} V_{cq_{1}}^{\ast} V_{uq_{2}}  \,
     \Big[ C_{1}({\mu}) \, O_{1}
         + C_{2}({\mu}) \, O_{2} \Big]
     + {\rm h.c.}
     \label{Hamiltonian-c},
     \end{equation}
     \begin{equation}
     O_{1} \, = \,
     [ \bar{q}_{1,{\alpha}}\, {\gamma}_{\mu}\, (1-{\gamma}_{5}) \, c_{\alpha} ] \
     [ \bar{u}_{\beta}\, {\gamma}^{\mu}\, (1-{\gamma}_{5}) \, q_{2,{\beta}} ]
     \label{c-o1},
     \end{equation}
     \begin{equation}
     O_{2} \, = \,
     [ \bar{q}_{1,{\alpha}}\, {\gamma}_{\mu}\, (1-{\gamma}_{5}) \, c_{\beta} ] \
     [ \bar{u}_{\beta}\, {\gamma}^{\mu}\, (1-{\gamma}_{5}) \, q_{2,{\alpha}} ]
     \label{c-o2},
     \end{equation}
     where $G_{F}$ ${\approx}$ $1.166\,{\times}\,10^{-5}\,{\rm GeV}^{-2}$
     \cite{PhysRevD.110.030001}
     is the Fermi coupling constant of the weak interactions;
     $C_{1,2}$ is the Wilson coefficient;
     the local operator $O_{1,2}$ denotes the four-quark
     point-like weak interactions;
     ${\alpha}$ and ${\beta}$ are the $SU(3)$ color indices;
     $V_{uq_{1}}$ and $V_{cq_{2}}$ are the
     quark-mixing Cabibbo–Kobayashi–Maskawa (CKM) matrix
     elements, with $q_{1,2}$ $=$ $d$ and $s$.
     Using the Wolfenstein parameterization, the CKM elements,
     up to ${\cal O}({\lambda}^{6})$, can be written as
     \cite{PhysRevD.110.030001},
     \begin{eqnarray}
     V_{ud} & = &
            1 - \frac{1}{2} {\lambda}^{2}
              - \frac{1}{8} {\lambda}^{4}
              +{\cal O}({\lambda}^{6}),
     \label{vud} \\
     V_{us} & = & {\lambda} + {\cal O}({\lambda}^{6}),
     \label{vus} \\
     V_{ub} & = & A {\lambda}^{3} ( {\rho} - i {\eta} ),
     \label{vub} \\
     V_{cd} & = &
              - {\lambda}
              + A^{2} {\lambda}^{5} \Big[ \frac{1}{2} -( {\rho} + i {\eta} ) \Big]
              + {\cal O}({\lambda}^{6})
     \label{vcd} \\
     V_{cs} & = &
            1 - \frac{1}{2} {\lambda}^{2}
              - \frac{1}{8} {\lambda}^{4}
              - \frac{1}{2} A^{2} {\lambda}^{4}
              +{\cal O}({\lambda}^{6}),
     \label{vcs} \\
     V_{cb} & = & A {\lambda}^{2} + {\cal O}({\lambda}^{6}).
     \label{vcb}
     \end{eqnarray}
     The latest numerical values of the Wolfenstein parameters
     given by PDG \cite{PhysRevD.110.030001} are,
     \begin{eqnarray}
     A & = & 0.826^{+0.016}_{-0.015}
     \label{Wolfenstein-a}, \\
    {\lambda}  & = & 0.22501 {\pm} 0.00068
     \label{Wolfenstein-lambda}, \\
     \bar{\rho} & = & 0.1591 {\pm} 0.0094
     \label{Wolfenstein-rho}, \\
     \bar{\eta} & = & 0.3523^{+0.0073}_{-0.0071}
     \label{Wolfenstein-eta}.
     \end{eqnarray}

     It should be pointed out that
     (1)
     both the penguin and annihilation contributions being
     proportional to the CKM factor
     $V_{ub}\,V_{cb}^{\ast}$ ${\sim}$ ${\cal O}({\lambda}^{5})$
     are highly suppressed relative to those from the operator
     $O_{1,2}$ being proportional to
     $V_{ud}\,V_{cs}^{\ast}$ ${\sim}$ ${\cal O}(1)$
     or $V_{ud}\,V_{cd}^{\ast}$ ${\sim}$ ${\cal O}({\lambda})$
     or $V_{us}\,V_{cs}^{\ast}$ ${\sim}$ ${\cal O}({\lambda})$
     or $V_{us}\,V_{cd}^{\ast}$ ${\sim}$ ${\cal O}({\lambda}^{2})$.
     Hence, the penguin and annihilation contributions to branching
     ratios can be safely neglected here for
     the moment.
     (2)
     The Wilson coefficients $C_{i}$ summarize the physics
     contributions from scale higher than ${\mu}$ and can be
     calculated with the perturbation theory.
     It is generally recognized that the Wilson coefficients $C_{i}$
     have nothing to do with the
     hadronization process of quarks, and so can be factorized
     from specific hadron processes, {\em i.e.},
     they are generally universal and process-independent.
     Their values including the next-to-leading order (NLO)
     corrections can be evaluated with the renormalization
     group equation \cite{RevModPhys.68.1125}.
     The numerical values of Wilson coefficients for the $c$ and
     $b$ quark decays can be found in numerous references, such as
     Refs. \cite{PhysRevD.77.074013,PhysRevD.77.114004,AHEP.2015.104378}.

     For the $B_{c}(2S)$ ${\to}$ ${\psi}(1S,2S)\,P$,
     ${\eta}_{c}(1S,2S)\,P$ decays induced by the $b$ ${\to}$ $c$
     quark transition, the expression of the low-energy effective
     Hamiltonian is written as \cite{PhysRevD.77.074013},
     \begin{equation}
    {\cal H}_{\rm eff}^{b} \, = \,
     \frac{G_{F}}{\sqrt{2}}
     \sum\limits_{q=d,s}  V_{cb} V_{uq}^{\ast} \,
     \Big[ C_{1}({\mu}) \, O_{1}^{\prime}
         + C_{2}({\mu}) \, O_{2}^{\prime} \Big]
     + {\rm h.c.}
     \label{Hamiltonian-b},
     \end{equation}
     \begin{equation}
     O_{1}^{\prime} \, = \,
     [ \bar{c}_{{\alpha}}\, {\gamma}_{\mu}\, (1-{\gamma}_{5}) \, b_{\alpha} ] \
     [ \bar{q}_{\beta}\, {\gamma}^{\mu}\, (1-{\gamma}_{5}) \, u_{{\beta}} ]
     \label{b-o1},
     \end{equation}
     \begin{equation}
     O_{2}^{\prime} \, = \,
     [ \bar{c}_{{\alpha}}\, {\gamma}_{\mu}\, (1-{\gamma}_{5}) \, b_{\beta} ] \
     [ \bar{q}_{\beta}\, {\gamma}^{\mu}\, (1-{\gamma}_{5}) \, u_{{\alpha}} ]
     \label{b-o2}.
     \end{equation}
     Note that because the valence quarks in both final states differ
     from each other, there is no contributions from penguin and
     annihilation in Eq.(\ref{Hamiltonian-b}).

     Taking the $B_{c}(2S)$ ${\to}$ ${\psi}\, {\pi}$ decay as an example,
     the decay amplitude can be written as,
     \begin{equation}
    {\cal A} \, = \,
    {\langle} \, {\psi}\, {\pi} \, {\vert} \,
    {\cal H}_{\rm eff} \, {\vert} \, B_{c}(2S) \, {\rangle}
     \, = \,
     \frac{G_{F}}{\sqrt{2}} \, V_{cb} \, V_{ud}^{\ast} \,
     \sum\limits_{i=1}^{2} \, C_{i} \,
    {\langle} \, {\psi}\, {\pi} \, {\vert}  \, O_{i}^{\prime} \, {\vert} \, B_{c}(2S) \, {\rangle}
     \label{decay-amplitude},
     \end{equation}
     where ${\langle} \, {\psi}\, {\pi} \, {\vert}  \, O_{i}^{\prime} \, {\vert} \, B_{c}(2S) \, {\rangle}$
     is usually called as the hadronic matrix element (HME).
     HME describes the transitions from the local effective operator
     at the quark level to the initial and final states at the hadron
     level.
     The participation of the strong interactions make it very
     complicated to calculate HME, which including the perturbative
     and nonperturbative contributions.
     To obtain the decay amplitudes, the remaining and most complex
     task is to accurately evaluate HMEs.

     \subsection{Hadronic matrix elements}
     \label{sec0202}
     The QCDF approach \cite{PhysRevLett.83.1914,NuclPhysB.591.313,
     NuclPhysB.606.245,PhysLettB.488.46,PhysLettB.509.263,PhysRevD.64.014036}
     is one of the recently-developed QCD-improved phenomenological
     models to deal with HME, based on the exclusive hard scattering
     approach \cite{PhysRevD.22.2157} and the power series expansion
     in the heavy quark limit.
     At the leading power of inverse heavy quark mass, the master
     QCDF formula for HME is generally expressed as a convolution
     integral form, incorporating both the perturbative (hard
     scattering kernel) and nonperturbative (hadronic wave
     function) parts;
     specifically, for the case with a heavy recoiled meson,
     HME is written as \cite{NuclPhysB.591.313},
     \begin{equation}
    {\langle} \, {\psi}\, {\pi} \, {\vert}  \, O_{i}^{\prime} \, {\vert} \, B_{c}(2S) \, {\rangle}
     \, = \, F^{B_{c}(2S){\to}{\psi}} {\int}_{0}^{1} dz \, T(z) \, {\phi}_{\pi}(z)
     \label{qcdf-hme},
     \end{equation}
     where both the transition form factor $F^{B_{c}(2S){\to}{\psi}}$
     and the light-cone distribution amplitude ${\phi}_{\pi}$ of
     the emitted pion are
     nonperturbative but universal parameters at the hadron level,
     and can be obtained from other nonperturbative method or
     extracted from data.
     The scattering kernel $T$ arising from hard gluon exchange
     among quarks are perturbatively calculable.
     By including the QCD radiative corrections to HME, part of the
     strong phases can be retrieved for $CP$ asymmetries, and the
     renormalization scale dependence of HME can be recuperated to
     cancel the scale dependence of the Wilson coefficients and
     obtain a physical decay amplitude.

     The decay amplitude in Eq.(\ref{decay-amplitude})
     with the QCDF approach can be written as,
     \begin{equation}
    {\cal A} \, = \, \sqrt{2}\, G_{F} \, V_{cb} \, V_{ud}^{\ast} \,
     A_{0}^{B_{c}(2S){\to}{\psi}} \, f_{\pi} \,
     m_{\psi} \, ({\epsilon}_{\psi}{\cdot}p_{\pi}) \, a_{1}
     \label{decay-amplitude-qcdf},
     \end{equation}
     where $A_{0}^{B_{c}(2S){\to}{\psi}}$ is the form factor,
     and its definition is given by Eq.(\ref{form-factor-1m}).
     The decay constant $f_{\pi}$ is defined
     by Eq.(\ref{p-a-decay-constant}), and it
     can also be obtained from the normalization condition,
     \begin{equation}
    {\int}_{0}^{1} dz \, {\phi}_{\pi}(z) \, = \, f_{\pi}
     \label{decay-constant},
     \end{equation}
     with the leading twist distribution amplitude
     \cite{JHEP.2006.05.004,JHEP.2007.03.069},
     \begin{equation}
    {\phi}_{\pi}(z) \, = \, f_{\pi} \, 6\, z\, \bar{z}
     \sum\limits_{n=0} a_{n}^{\pi} \, C_{n}^{3/2}(z-\bar{z})
     \label{decay-constant},
     \end{equation}
     where $z$ denotes the longitudinal momentum fractions
     of valence quark; $\bar{z}$ $=$ $1$ $-$ $z$;
     the expansion coefficient coefficient $a_{n}^{\pi}$ of
     the Gegenbauer polynomials $C_{n}^{3/2}$ is usually called
     as the Gegenbauer moment, which is a nonperturbative shape
     parameter.
     The QCDF coefficients $a_{i}$ including the NLO QCD
     radiative contributions are written as,
     \begin{equation}
     a_{1}  \, = \,
       C_{1}^{\rm NLO}
     + \frac{1}{N_{c}}\, C_{2}^{\rm NLO}
     + \frac{{\alpha}_{s}}{4{\pi}}\, \frac{C_{F}}{N_{c}}\, C_{2}^{\rm LO}\, V
     \label{qcdf-a1},
     \end{equation}
     \begin{equation}
     a_{2}  \, = \,
       C_{2}^{\rm NLO}
     + \frac{1}{N_{c}}\, C_{1}^{\rm NLO}
     + \frac{{\alpha}_{s}}{4{\pi}}\, \frac{C_{F}}{N_{c}}\, C_{1}^{\rm LO}\, V
     \label{qcdf-a2},
     \end{equation}
     where the analytic expression of vertex correction
     factor $V$ will no longer be displayed here.
     The details of the vertex factor $V$ and decay amplitudes have
     already been addressed in Ref. \cite{PhysRevD.77.074013} for the
     $B_{c}(2S)$ ${\to}$ ${\psi}\,P$, ${\eta}_{c}\,P$ decays,
     and Ref. \cite{AHEP.2015.104378} for the $B_{c}(2S)$
     ${\to}$ $B\,P$, $B\,V$ decays.
     It has been proven that the NLO QCDF coefficients
     $a_{i}$ are independent of the renormalization scale at the
     order of ${\alpha}_{s}$ \cite{PhysRevD.64.014036,PhysRevD.77.074013}.
     The QCDF coefficients $a_{i}$ beyond NLO can be found
     in Ref. \cite{JHEP.2016.09.112} for the $b$ ${\to}$
     $c$ transition, and Ref. \cite{NuclPhysB.832.109} for
     the $b$ ${\to}$ $q$ transition.
     As stated in Ref. \cite{NuclPhysB.832.109}, the coefficients
     $a_{1}$ is rather stable against radiative corrections.
     For a rough estimate to branching ratio for the $B_{c}(2S)$
     weak decay, the NLO QCDF coefficients $a_{i}$ are basically
     sufficient for the requirements.

     \subsection{Decay constants and form factors}
     \label{sec0203}
     In the above QCDF formula for HME, Eq.(\ref{qcdf-hme}),
     at the leading order in ${\alpha}_{s}$ expansion,
     the scattering kernel $T(z)$ $=$ $1$.
     With the normalization condition Eq.(\ref{decay-constant}),
     HME are parameterized as the product
     of form factors and decay constants.
     One will return to the simple ``naive factorization'' (NF)
     scenario, {\em i.e.}, HME in Eq.(\ref{qcdf-hme}) can be
     divided into two HMEs of color-singlet current operators,
     for example,
     \begin{equation}
    {\langle} \, {\psi}\, {\pi} \, {\vert}  \, O_{1}^{\prime} \, {\vert} \, B_{c}(2S) \, {\rangle}
     \, = \,
    {\langle} \, {\pi} \, {\vert}  \, \bar{d}_{\beta}\,
    {\gamma}^{\mu}\, (1-{\gamma}_{5}) \, u_{{\beta}} \,
    {\vert} \, 0 \, {\rangle} \
    {\langle} \, {\psi}\, {\vert}  \, \bar{c}_{{\alpha}}\,
    {\gamma}_{\mu}\, (1-{\gamma}_{5}) \, b_{\alpha} \,
    {\vert} \, B_{c}(2S) \, {\rangle}
     \label{nf-hme}.
     \end{equation}

     HME for creating a meson from the vacuum
     is parameterized by,
     \begin{equation}
    {\langle} \, P(p) \, {\vert} \, V_{\mu} \,
    {\vert} \, 0 \, {\rangle}
     \, = \, 0
     \label{p-v-decay-constant},
     \end{equation}
     \begin{equation}
    {\langle} \, P(p) \, {\vert} \, A_{\mu} \,
    {\vert} \, 0 \, {\rangle}
     \, = \, -i \, f_{P}\, p_{\mu}
     \label{p-a-decay-constant},
     \end{equation}
     \begin{equation}
    {\langle} \, V(p,{\epsilon}) \, {\vert} \, V_{\mu} \,
    {\vert} \, 0 \, {\rangle}
     \, = \,
     f_{V} \, m_{V} \, {\epsilon}_{V,{\mu}}^{\ast}
     \label{v-v-decay-constant},
     \end{equation}
     \begin{equation}
    {\langle} \, V(p,{\epsilon}) \, {\vert} \, A_{\mu} \,
    {\vert} \, 0 \, {\rangle}
     \, = \, 0
     \label{v-a-decay-constant},
     \end{equation}
     where $f_{P}$ and $f_{V}$ are the decay constants
     of pseudoscalar $P$ and vector $V$ mesons, respectively;
     $m_{V}$ and ${\epsilon}_{V}$ denote the mass and
     polarization of vector meson, respectively.

     For the mixing of physical pseudoscalar ${\eta}$ and
     ${\eta}^{\prime}$ meson, we adopt the quark-flavor
     basis description proposed in Ref. \cite{PhysRevD.58.114006},
     and neglect the contributions from possible gluonium and
     $c\bar{c}$ compositions, {\em i.e.},
     \begin{equation}
     \Bigg( \begin{array}{c}
    {\eta} \\ {\eta}^{\prime}
     \end{array} \Bigg) \, = \,
     \Bigg( \begin{array}{cc}
    {\cos}{\phi} & -{\sin}{\phi} \\
    {\sin}{\phi} & ~{\cos}{\phi}
     \end{array} \Bigg) \,
     \Bigg( \begin{array}{c}
    {\eta}_{q} \\ {\eta}_{s}
     \end{array} \Bigg)
     \label{etaq-etas-mixing},
     \end{equation}
     where ${\eta}_{q}$ $=$ $(u\bar{u}+d\bar{d})/{\sqrt{2}}$
     and ${\eta}_{s}$ $=$ $s\bar{s}$;
     the mixing angle ${\phi}$ $=$ $(39.3{\pm}1.0)^{\circ}$
     \cite{PhysRevD.58.114006}.

     Here, we will adopt a convention for hadronic transition
     form factors defined in Ref. \cite{ZPhysC.29.637}.
     \begin{eqnarray} &   &
    {\langle} \, X(p_{2})\, {\vert}  \,
     \bar{q}_{\alpha} \, {\gamma}^{\mu} \, (1-{\gamma}_{5}) \,  b_{\alpha} \,
    {\vert} \, B_{c}(2S)(p_{1}) \, {\rangle}
     \nonumber \\ & = &
     \frac{ m_{B_{c}(2S)}^{2}-m_{X}^{2} }{q^{2}} \, q^{\mu} \, F_{0}^{B_{c}(2S){\to}X}(q^{2})
   + \Big( p_{1}^{\mu} + p_{2}^{\mu}
   - \frac{ m_{B_{c}(2S)}^{2}-m_{X}^{2} }{q^{2}} \, q^{\mu}
     \Big) \, F_{1}^{B_{c}(2S){\to}X}(q^{2})
     \label{form-factor-0m},
     \end{eqnarray}
     \begin{eqnarray} &   &
    {\langle} \, {\psi} (p_{2},{\epsilon})\, {\vert}  \,
     \bar{q}_{\alpha} \, {\gamma}^{\mu} \, (1-{\gamma}_{5}) \,  b_{\alpha} \,
    {\vert} \, B_{c}(2S)(p_{1}) \, {\rangle}
     \nonumber \\ & = &
    {\epsilon}_{{\mu}{\nu}{\alpha}{\beta}} \,
    {\epsilon}_{\psi}^{{\ast}{\nu}}\,
     p_{1}^{\alpha}\, p_{2}^{\beta}\,
     \frac{ 2 \, V^{B_{c}(2S){\to}{\psi}}(q^{2}) }
          { m_{B_{c}(2S)}+m_{\psi} }
     + i\, \frac{ 2\, m_{\psi}\, ( {\epsilon}_{\psi}^{\ast}{\cdot}q) }{ q^{2} }\,
        q_{\mu}\, A_{0}^{B_{c}(2S){\to}{\psi}}(q^{2})
     \nonumber \\ & + &
       i\, {\epsilon}_{{\psi},{\mu}}^{\ast} \,
       ( m_{B_{c}(2S)}+m_{{\psi}} )\, A_{1}^{B_{c}(2S){\to}{\psi}}(q^{2})
     - i\, \frac{ ({\epsilon}_{{\psi}}^{\ast}{\cdot}q) }
                { m_{B_{c}(2S)}+m_{\psi} }\,
       ( p_{1} + p_{2} )_{\mu}\, A_{2}^{B_{c}(2S){\to}{\psi}}(q^{2})
     \nonumber \\ & - &
       i\, \frac{ 2\, m_{\psi}\, ({\epsilon}_{\psi,{\mu}}^{\ast}{\cdot}q) }{ q^{2} }\,
        q_{\mu}\, A_{3}^{B_{c}(2S){\to}{\psi}}(q^{2})
     \label{form-factor-1m},
     \end{eqnarray}
     where
     $X$ denotes the pesudoscalar $B_{u,d,s}$,
     ${\eta}_{c}(1S)$ and ${\eta}_{c}(2S)$ mesons;
     ${\psi}$ denotes the vector ${\psi}(1S)$ and ${\psi}(2S)$ mesons;
     the momentum transfer $q$ $=$ $p_{1}$ $-$ $p_{2}$;
     ${\epsilon}_{\psi}^{\ast}$ denotes the polarization vector of the ${\psi}$ meson;
     $F_{0,1}^{B_{c}(2S){\to}X}$,
     $V^{B_{c}(2S){\to}{\psi}}$ and
     $A_{0,1,2,3}^{B_{c}(2S){\to}{\psi}}$
     are the transition form factors.
     At the large recoil limit, $q^{2}$ $=$ $0$, there is
     \begin{equation}
     F_{0}^{B_{c}{\to}X}(0) \, = \,
     F_{1}^{B_{c}{\to}X}(0),
     \label{form-factor-f0f1}
     \end{equation}
     \begin{equation}
     A_{0}^{B_{c}{\to}{\psi}}(0) \, = \,
     A_{3}^{B_{c}{\to}{\psi}}(0).
     \label{form-factor-a0a3}
     \end{equation}

     Currently, there is no theoretical calculation for these form
     factors. We will use the Wirbel-Stech-Bauer approach \cite{ZPhysC.29.637}
     to give a rough estimation of the form factor.
     The form factors at the pole $q^{2}$ $=$ $0$ are defined as
     the convolution of wave functions of both the initial and
     the recoiled mesons \cite{ZPhysC.29.637}, {\it i.e.},
     \begin{equation}
     F_{0}^{B_{c}(2S){\to}X}(0) \, = \,
    {\int}d \vec{k}_{\perp} {\int}_{0}^{1}dx \,
    {\Phi}_{X}(\vec{k}_{\perp},x,0,0) \,
    {\Phi}_{B_{c}(2S)}(\vec{k}_{\perp},x,0,0)
     \label{form-f0-wave},
     \end{equation}
     \begin{equation}
     A_{0}^{B_{c}(2S){\to}{\psi}}(0) \, = \,
    {\int}d \vec{k}_{\perp} {\int}_{0}^{1}dx \,
    {\Phi}_{\psi}(\vec{k}_{\perp},x,1,0)\,
    {\sigma}_{z}\, {\Phi}_{B_{c}(2S)}(\vec{k}_{\perp},x,0,0)
     \label{form-a0-wave},
     \end{equation}
     where ${\sigma}_{z}$ is the Pauli matrix acting on the
     spin indices of the decaying quark;
     the variables $\vec{k}_{\perp}$, $x$, $J$, $J_{z}$ of the
     wave function ${\Phi}(\vec{k}_{\perp},x,J,J_{z})$
     denote respectively the transverse momentum and the
     fraction of the longitudinal momentum carried by
     the nonspectator quark, the total spin and
     its $z$-component.
     Both the $B_{c}(2S)$ meson and charmonium consist of two
     heavy flavors.
     The motion of the valence quarks in these mesons should be
     nonrelativistic.
     For the $B_{c}(2S)$ and charmonium bound states,
     the radial solution of the Sch\"{o}dinger equation with
     an isotropic harmonic oscillator potential by separation
     of the angular variables, after the Fourier transformation
     from coordinate space to momentum domain,
     is written as
     \cite{PhysLettB.751.171,PhysLettB.752.322,JPG.42.105005,
     AHEP.2016.5071671,PhysRevD.95.036024},
     \begin{equation}
    {\Phi}_{B_{c}(2S)}(\vec{k}_{\perp},x) \, = \, A  \,
     \Big\{ \frac{ \vec{k}_{\perp}^{2} + \bar{x}\, m_{c}^{2} + x\, m_{b}^{2} }
                       { 6\, {\alpha}_{1}^{2}\, x\, \bar{x} } - 1 \Big\} \,
    {\exp} \Big\{ - \frac{ \vec{k}_{\perp}^{2} + \bar{x}\, m_{c}^{2} + x\, m_{b}^{2} }
                       { 8\, {\alpha}_{1}^{2}\, x\, \bar{x} } \Big\}
     \label{wave-bc-2s},
     \end{equation}
     \begin{equation}
    {\Phi}_{{\eta}_{c}(1S)}(\vec{k}_{\perp},x) \, = \,
    {\Phi}_{{\psi}(1S)}(\vec{k}_{\perp},x) \, = \, B \,
    {\exp} \Big\{ - \frac{ \vec{k}_{\perp}^{2} + m_{c}^{2} }
                       { 8\, {\alpha}_{2}^{2}\, x\, \bar{x} } \Big\}
     \label{wave-cc-1s},
     \end{equation}
     \begin{equation}
    {\Phi}_{{\eta}_{c}(2S)}(\vec{k}_{\perp},x) \, = \,
    {\Phi}_{{\psi}(2S)}(\vec{k}_{\perp},x) \, = \, C
     \Big\{ \frac{ \vec{k}_{\perp}^{2} + m_{c}^{2} }
                       { 6\, {\alpha}_{2}^{2}\, x\, \bar{x} } -1 \Big\} \,
    {\exp} \Big\{ - \frac{ \vec{k}_{\perp}^{2} + m_{c}^{2} }
                       { 8\, {\alpha}_{2}^{2}\, x\, \bar{x} } \Big\}
     \label{wave-cc-2s},
     \end{equation}
     where
     the mean value of square of transverse momentum is
     ${\alpha}_{1}^2$ $=$ ${\mu}\,{\omega}$ with the
     reduced mass ${\mu}$ $=$ $\frac{m_{b}\,m_{c}}{m_{b}+m_{c}}$
     and the characteristic frequency ${\omega}$ ${\approx}$
     $0.50$ ${\pm}$ $0.05$ GeV \cite{PhysRevD.89.114019,
     PhysRevD.49.5845,PhysRevD.60.074006,PhysRevD.70.054017,
     PhysRevD.100.096002,PhysRevD.106.054009,
     PhysRevD.51.3613,PhysRevD.52.5229,PhysRevD.53.312,
     PhysRevD.67.014027,EPJC.71.1825,PhysRevD.95.054016,
     EPJC.78.592,PhysRevD.99.054025,PhysRevD.99.096020,
     JHEP.2205.006,EPJC.83.1080};
     and ${\alpha}_{2}$ $=$ $m_{c}\,{\alpha}_{s}$
     for the charmonium \cite{AHEP.2016.5071671,PhysRevD.95.036024}.
     The parameters of $A$, $B$, $C$ are the normalization coefficients.
     For the $B_{u,d,s}$ meson, we will adopt the Gaussian type
     proposed in \cite{PhysRevD.63.054008}, after the
     Fourier transformation,
     \begin{equation}
    {\Phi}_{B_{u,d,s}}(\vec{k}_{\perp},x) \, = \, D \, x^{2}\, \bar{x}^{2}\,
    {\exp} \Big\{ - \frac{ \vec{k}_{\perp}^{2} + x^{2}\, m_{B}^{2} }
                       { 2\, {\alpha}_{3}^{2} } \Big\},
     \label{wave-bq},
     \end{equation}
     where $D$ is the normalization constant;
     ${\alpha}_{3}$ $=$ $0.45$ ${\pm}$ $0.05$ GeV for $B_{u,d}$ meson
     and $0.55$ ${\pm}$ $0.05$ GeV for $B_{s}$ meson
     \cite{PhysRevD.89.114019,PhysRevD.103.056006,CPC.46.083103,EPJC.85.544}.

     \begin{table}[b]
     \caption{The numerical values of the input parameters,
          where their central values are regarded as the default
          inputs unless otherwise specified.}
     \label{tab:input}
     \begin{ruledtabular}
     \begin{tabular}{llll}
     \multicolumn{4}{c}{ Mass of particle (in units of MeV) \cite{PhysRevD.110.030001} }  \\ \hline
       $m_{B_{u}}$ $=$ $5279.41(7)$
     & $m_{{\eta}_{c}(1S)}$ $=$ $2984.1(4)$
     & $m_{{\pi}^{\pm}}$ $=$ $139.57$
     & $m_{\rho}$ $=$ $775.26(23)$
     \\
       $m_{B_{d}}$ $=$ $5279.72(8)$
     & $m_{{\psi}(1S)}$ $=$ $3096.900(6)$
     & $m_{{\pi}^{0}}$ $=$ $134.98$
     & $m_{\omega}$ $=$ $782.66(13)$
     \\
       $m_{B_{s}}$ $=$ $5366.93(10)$
     & $m_{{\eta}_{c}(2S)}$ $=$ $3637.7(9)$
     & $m_{{\eta}}$ $=$ $547.862(17)$
     & $m_{\phi}$ $=$ $1019.461(16)$
     \\
       $m_{B_{c}(2S)}$ $=$ $6871.2(1.0)$
     & $m_{{\psi}(2S)}$ $=$ $3686.097(11)$
     & $m_{{\eta}^{\prime}}$ $=$ $957.78(6)$
     & $m_{K^{{\ast}{\pm}}}$ $=$ $895.5(8)$
     \\
       $m_{b}$ $=$ $4183(7)$
     & $m_{K^{\pm}}$ $=$ $493.677(15)$
     & $m_{K^{0}}$   $=$ $497.611(13)$
     & $m_{K^{{\ast}0}}$ $=$ $895.55(20)$
     \\ \hline
     \multicolumn{4}{c}{Decay constant (in units of MeV)}  \\ \hline
       $f_{\pi}$ $=$ $130.2(1.2)$ \cite{PhysRevD.110.030001}
     & $f_{K}$   $=$ $155.7(3)$   \cite{PhysRevD.110.030001}
     & $f_{{\eta}_{q}}$ $=$ $1.07(2)\,f_{\pi}$ \cite{PhysRevD.58.114006}
     & $f_{{\eta}_{s}}$ $=$ $1.34(6)\,f_{\pi}$ \cite{PhysRevD.58.114006}
     \\
       $f_{\rho}$     $=$ $216(3)$ \cite{JHEP.2007.03.069}
     & $f_{\omega}$   $=$ $187(5)$ \cite{JHEP.2007.03.069}
     & $f_{\phi}$     $=$ $215(5)$ \cite{JHEP.2007.03.069}
     & $f_{K^{\ast}}$ $=$ $220(5)$ \cite{JHEP.2007.03.069}
     \\ \hline
     \multicolumn{4}{c}{Gegenbauer moments at the scale ${\mu}$ $=$ 1 GeV
         \cite{JHEP.2006.05.004,JHEP.2007.03.069}}  \\ \hline
       $a_{1}^{\bar{K}}$ $=$ $-a_{1}^{K}$ $=$ $0.06(3)$
     & $a_{2}^{\bar{K}}$ $=$ $ a_{2}^{K}$ $=$ $0.25(15)$
     & $a_{2}^{\pi}$ $=$ $0.25(15)$
     \\
       $a_{1}^{\bar{K}^{\ast}}$ $=$ $-a_{1}^{K^{\ast}}$ $=$  $0.03(2)$
     & $a_{2}^{\bar{K}^{\ast}}$ $=$ $ a_{2}^{K^{\ast}}$ $=$  $0.11(9)$
     & $a_{2}^{\rho}$ $=$ $a_{2}^{\omega}$ $=$ $0.15(7)$
     & $a_{2}^{\phi}$ $=$ $0.18(8)$
     \end{tabular}
     \end{ruledtabular}
     \end{table}

     According to the above definition and conventions, and using the input
     parameters listed in Table \ref{tab:input}, we obtain the
     numerical values of form factors at the pole $q^{2}$ $=$ $0$,
     \begin{equation}
     F_{0}^{B_{c}(2S){\to}{B_{u,d}}}(0) \, = \, 0.397
     \label{f0-bq},
     \end{equation}
     \begin{equation}
     F_{0}^{B_{c}(2S){\to}{B_{s}}}(0) \, = \, 0.426
     \label{f0-bs},
     \end{equation}
     \begin{equation}
     F_{0}^{B_{c}(2S){\to}{\eta}_{c}(1S)}(0) \, = \,
     A_{0}^{B_{c}(2S){\to}    {\psi}(1S)}(0) \, = \, 0.270
     \label{f0-etac1s},
     \end{equation}
     \begin{equation}
     F_{0}^{B_{c}(2S){\to}{\eta}_{c}(2S)}(0) \, = \,
     A_{0}^{B_{c}(2S){\to}    {\psi}(2S)}(0) \, = \, 0.260
     \label{f0-etac2s},
     \end{equation}
     which will be used to provide an order-of-magnitude
     estimation of the branching ratio of the $B_{c}(2S)$ meson
     weak decay.

     \section{Numerical results and discussion}
     \label{sec03}
     In the rest frame of $B_{c}(2S)$, branching ratio
     for the nonleptonic two-body $B_{c}(2S)$ weak decays
     is written as,
     \begin{equation}
    {\cal B}r \, = \, \frac{ p_{\rm cm} }{ 8\, {\pi}\, m_{B_{c}(2S)}^{2}\, {\Gamma}_{B_{c}(2S)} } \,
    {\vert} {\cal A} {\vert}^{2}
     \label{eq:br},
     \end{equation}
     where $p_{\rm cm}$ is the common momentum of final particles;
     ${\Gamma}_{B_{c}(2S)}$ is the total width of the $B_{c}(2S)$ meson.
     The experimental data on ${\Gamma}_{B_{c}(2S)}$ is unavailable at present.
     According to the numbers in Table \ref{width-bc2s}
     obtained with various theoretical models, we will
     use ${\Gamma}_{B_{c}(2S)}$ $=$ $55$ keV for the moment.
     The similar amplitudes ${\cal A}$ for the $B_{c}(2S)$
     ${\to}$ $B\,P$, $B\,V$ decays can be found in
     Refs. \cite{PhysRevD.77.114004,AHEP.2015.104378},
     and the $B_{c}(2S)$ ${\to}$ ${\psi}\,P$, ${\eta}_{c}\,P$
     decays in Ref. \cite{PhysRevD.77.074013},
     with the replacement of symbol $B_{c}$ by $B_{c}(2S)$.

     \begin{table}[t]
     \caption{The numerical results on $CP$-averaged branching
      ratios for the $B_{c}(2S)$ ${\to}$ $BP$, $BV$ decays,
      where the theoretical uncertainties come from the CKM parameters,
      the renormalization scale ${\mu}$ $=$ $(1{\pm}0.2)\,m_{c}$,
      decay constants, and Gegenbauer moments, respectively.}
     \label{br:c-decay}
     \begin{ruledtabular}
     \begin{tabular}{lcccc}
       \multicolumn{1}{c}{final}
     & \multicolumn{2}{c}{decay amplitude (${\cal A}$)}
     & &
      \\ \cline{2-3}
       \multicolumn{1}{c}{states}
     & coefficient & CKM factor
     & case & branching ratio (${\cal B}r$)
      \\ \hline
       $B_{s}^{0} \, {\pi}^{+}$
     & $a_{1}$
     & $V_{cs}^{\ast}\,V_{ud}$ ${\sim}$ ${\cal O}(1)$
     & I-a
     & $( 1.93^{+0.00+0.18+0.01}_{
                -0.00-0.10-0.01} )\, {\times}\, 10^{-9}$
     \\
       $B_{s}^{0} \, {\rho}^{+}$
     & $a_{1}$
     & $V_{cs}^{\ast}\,V_{ud}$ ${\sim}$ ${\cal O}(1)$
     & I-a
     & $ ( 3.34^{+0.00+0.31+0.10}_{
                 -0.00-0.17-0.10} )\, {\times}\, 10^{-9}$
     \\
       $B_{s}^{0} \, K^{+}$
     & $a_{1}$
     & $V_{cs}^{\ast}\,V_{us}$ ${\sim}$ ${\cal O}({\lambda})$
     & I-b
     & $ ( 1.40^{+0.01+0.13+0.02}_{
                 -0.01-0.07-0.02} )\, {\times}\, 10^{-10}$
     \\
       $B_{s}^{0} \, K^{{\ast}+}$
     & $a_{1}$
     & $V_{cs}^{\ast}\,V_{us}$ ${\sim}$ ${\cal O}({\lambda})$
     & I-b
     & $ ( 1.53^{+0.01+0.14+0.07}_{
                 -0.01-0.08-0.07} )\, {\times}\, 10^{-10}$
     \\
       $B_{d}^{0} \, {\pi}^{+}$
     & $a_{1}$
     & $V_{cd}^{\ast}\,V_{ud}$ ${\sim}$ ${\cal O}({\lambda})$
     & I-b
     & $( 1.04^{+0.01+0.10+0.01}_{
                -0.01-0.05-0.01} )\, {\times}\, 10^{-10}$
     \\
       $B_{d}^{0} \, {\rho}^{+}$
     & $a_{1}$
     & $V_{cd}^{\ast}\,V_{ud}$ ${\sim}$ ${\cal O}({\lambda})$
     & I-b
     & $ ( 1.90^{+0.01+0.18+0.06}_{
                 -0.01-0.10-0.05} )\, {\times}\, 10^{-10}$
     \\
       $B_{d}^{0} \, K^{+}$
     & $a_{1}$
     & $V_{cd}^{\ast}\,V_{us}$ ${\sim}$ ${\cal O}({\lambda}^{2})$
     & I-c
     & $ ( 7.60^{+0.09+0.72+0.10}_{
                 -0.09-0.40-0.10} )\, {\times}\, 10^{-12}$
     \\
       $B_{d}^{0} \, K^{{\ast}+}$
     & $a_{1}$
     & $V_{cd}^{\ast}\,V_{us}$ ${\sim}$ ${\cal O}({\lambda}^{2})$
     & I-c
     & $ ( 8.97^{+0.11+0.84+0.43}_{
                 -0.11-0.47-0.42} )\, {\times}\, 10^{-12}$
     \\ \hline
       $B_{u}^{+} \, \overline{K}^{0}$
     & $a_{2}$
     & $V_{cs}^{\ast}\,V_{ud}$ ${\sim}$ ${\cal O}(1)$
     & II-a
     & $ ( 2.85^{+0.00+1.60+0.07}_{
                 -0.00-0.79-0.07} )\, {\times}\, 10^{-10}$
     \\
       $B_{u}^{+} \, \overline{K}^{{\ast}0}$
     & $a_{2}$
     & $V_{cs}^{\ast}\,V_{ud}$ ${\sim}$ ${\cal O}(1)$
     & II-a
     & $ ( 3.33^{+0.00+1.87+0.19}_{
                 -0.00-0.92-0.18} )\, {\times}\, 10^{-10}$
     \\
       $B_{u}^{+} \, {\pi}^{0}$
     & $a_{2}$
     & $V_{cd}^{\ast}\,V_{ud}$ ${\sim}$ ${\cal O}({\lambda})$
     & II-b
     & $ ( 5.61^{+0.03+3.15+0.09}_{
                 -0.03-1.55-0.09} )\, {\times}\, 10^{-12}$
     \\
       $B_{u}^{+} \, {\rho}^{0}$
     & $a_{2}$
     & $V_{cd}^{\ast}\,V_{ud}$ ${\sim}$ ${\cal O}({\lambda})$
     & II-b
     & $ ( 1.02^{+0.01+0.57+0.04}_{
                 -0.01-0.28-0.03} )\, {\times}\, 10^{-11}$
     \\
       $B_{u}^{+} \, {\omega}$
     & $a_{2}$
     & $V_{cd}^{\ast}\,V_{ud}$ ${\sim}$ ${\cal O}({\lambda})$
     & II-b
     & $ ( 7.55^{+0.04+4.24+0.46}_{
                 -0.04-2.09-0.44} )\, {\times}\, 10^{-12}$
     \\
       $B_{u}^{+} \, {\eta}$
     & $a_{2}$
     & $V_{cs}^{\ast}\,V_{us}$,  $V_{cd}^{\ast}\,V_{ud}$
     & II-b
     & $ ( 2.17^{+0.01+1.22+0.19}_{
                 -0.01-0.60-0.18} )\, {\times}\, 10^{-11}$
     \\
       $B_{u}^{+} \, {\eta}^{\prime}$
     & $a_{2}$
     & $V_{cs}^{\ast}\,V_{us}$,  $V_{cd}^{\ast}\,V_{ud}$
     & II-b
     & $ ( 2.79^{+0.02+1.56+0.64}_{
                 -0.02-0.77-0.56} )\, {\times}\, 10^{-12}$
     \\
       $B_{u}^{+} \, {\phi}$
     & $a_{2}$
     & $V_{cs}^{\ast}\,V_{us}$ ${\sim}$ ${\cal O}({\lambda})$
     & II-b
     & $ ( 1.37^{+0.01+0.77+0.00}_{
                 -0.01-0.38-0.00} )\, {\times}\, 10^{-11}$
     \\
       $B_{u}^{+} \, K^{0}$
     & $a_{2}$
     & $V_{cd}^{\ast}\,V_{us}$ ${\sim}$ ${\cal O}({\lambda}^{2})$
     & II-c
     & $ ( 8.21^{+0.10+4.61+0.21}_{
                 -0.10-2.27-0.21} )\, {\times}\, 10^{-13}$
     \\
       $B_{u}^{+} \, K^{{\ast}0}$
     & $a_{2}$
     & $V_{cd}^{\ast}\,V_{us}$ ${\sim}$ ${\cal O}({\lambda}^{2})$
     & II-c
     & $ ( 9.53^{+0.12+5.35+0.54}_{
                 -0.12-2.63-0.52} )\, {\times}\, 10^{-13}$
     \end{tabular}
     \end{ruledtabular}
     \end{table}
     \begin{table}[h]
     \caption{The numerical results on $CP$-averaged branching
      ratios for the $B_{c}(2S)$ ${\to}$ ${\eta}_{c}(1S,2S)\,P$,
      ${\psi}(1S,2S)\,P$ decays,
      where the theoretical uncertainties come from the CKM parameters,
      the renormalization scale ${\mu}$ $=$ $(1{\pm}0.2)\,m_{b}$,
      decay constants, and Gegenbauer moments, respectively.}
     \label{br:b-decay}
     \begin{ruledtabular}
     \begin{tabular}{lcccc}
       \multicolumn{1}{c}{final}
     & \multicolumn{2}{c}{decay amplitude (${\cal A}$)}
     & &
      \\ \cline{2-3}
       \multicolumn{1}{c}{states}
     & coefficient & CKM factor
     & case & branching ratio (${\cal B}r$)
      \\ \hline
       ${\eta}_{c}(1S)\, {\pi}^{+}$
     & $a_{1}$
     & $V_{cb}\,V_{ud}^{\ast}$ ${\sim}$ ${\cal O}({\lambda}^{2})$
     & I-c
     & $( 9.90^{+0.48+0.14+0.03}_{
                 -0.47-0.11-0.03} )\, {\times}\, 10^{-12}$
     \\
       ${\eta}_{c}(2S)\, {\pi}^{+}$
     & $a_{1}$
     & $V_{cb}\,V_{ud}^{\ast}$ ${\sim}$ ${\cal O}({\lambda}^{2})$
     & I-c
     & $( 6.40^{+0.31+0.10+0.02}_{
                -0.30-0.07-0.02} )\, {\times}\, 10^{-12}$
     \\
       ${\psi}(1S)\, {\pi}^{+}$
     & $a_{1}$
     & $V_{cb}\,V_{ud}^{\ast}$ ${\sim}$ ${\cal O}({\lambda}^{2})$
     & I-c
     & $( 9.36^{+0.46+0.14+0.03}_{
                -0.44-0.11-0.03} )\, {\times}\, 10^{-12}$
     \\
       ${\psi}(2S)\, {\pi}^{+}$
     & $a_{1}$
     & $V_{cb}\,V_{ud}^{\ast}$ ${\sim}$ ${\cal O}({\lambda}^{2})$
     & I-c
     & $( 6.19^{+0.30+0.09+0.02}_{
                -0.29-0.07-0.02} )\, {\times}\, 10^{-12}$
     \\
       ${\eta}_{c}(1S)\, K^{+}$
     & $a_{1}$
     & $V_{cb}\,V_{us}^{\ast}$ ${\sim}$ ${\cal O}({\lambda}^{3})$
     & I-d
     & $( 7.49^{+0.42+0.11+0.07}_{
                -0.40-0.09-0.07} )\, {\times}\, 10^{-13}$
     \\
       ${\eta}_{c}(2S)\, K^{+}$
     & $a_{1}$
     & $V_{cb}\,V_{us}^{\ast}$ ${\sim}$ ${\cal O}({\lambda}^{3})$
     & I-d
     & $( 4.83^{+0.27+0.07+0.04}_{
                -0.26-0.06-0.04} )\, {\times}\, 10^{-13}$
     \\
       ${\psi}(1S)\, K^{+}$
     & $a_{1}$
     & $V_{cb}\,V_{us}^{\ast}$ ${\sim}$ ${\cal O}({\lambda}^{3})$
     & I-d
     & $( 6.95^{+0.39+0.11+0.06}_{
                -0.37-0.08-0.06} )\, {\times}\, 10^{-13}$
     \\
       ${\psi}(2S)\, K^{+}$
     & $a_{1}$
     & $V_{cb}\,V_{us}^{\ast}$ ${\sim}$ ${\cal O}({\lambda}^{3})$
     & I-d
     & $( 4.56^{+0.25+0.07+0.04}_{
                -0.24-0.05-0.04}  )\, {\times}\, 10^{-13}$
     \end{tabular}
     \end{ruledtabular}
     \end{table}

     The numerical results on branching ratios for the $B_{c}(2S)$
     ${\to}$ $B\,P$, $B\,V$ decays are displayed in Table
     \ref{br:c-decay}, and the $B_{c}(2S)$ ${\to}$ ${\eta}_{c}(1S,2S)\,P$,
     ${\psi}(1S,2S)\,P$ decays in Table \ref{br:b-decay}.
     Our comments are as follows.

     (1)
     According to the QCDF coefficients $a_{1,2}$ and the CKM
     factors of the decay amplitudes, there is a clear hierarchy
     among the branching ratios for the $B_{c}(2S)$ weak decays,
     which can be categorized into I-a, I-b, I-c, I-d, II-a, II-b,
     II-c cases listed in Table \ref{br:c-decay} and \ref{br:b-decay},
     {\it i.e.},
     \begin{equation}
    {\cal B}r \, {\sim} \, \left\{
     \begin{array}{lcl}
    {\cal O}(10^{-9}),  & ~ & \text{for the I-a ~~~~~~ case}; \\
    {\cal O}(10^{-10}), & ~ & \text{for the I-b, II-a case}; \\
    {\cal O}(10^{-11} {\sim} 10^{-12}), &  & \text{for the ~~~~~ II-b case}; \\
    {\cal O}(10^{-12}), &  & \text{for the I-c ~~~~~~ case}; \\
    {\cal O}(10^{-13}), &  & \text{for the I-d, ~\!\!\! II-c case}.
     \end{array} \right.
     \label{br-hierarchy}
     \end{equation}

     (2)
     The $B_{c}(2S)$ ${\to}$ $B_{s}\,{\pi}$, $B_{s}\,{\rho}$ decays
     belonging to the I-a case have branching ratio ${\sim}$
     ${\cal O}(10^{-9})$.
     The $B_{c}(2S)$ ${\to}$ $B_{s}\,K^{(\ast)}$, $B_{d}\,{\pi}$,
     $B_{d}\,{\rho}$ decays belonging to the I-b case and the
     $B_{c}(2S)$ ${\to}$ $B_{u}\,\overline{K}^{(\ast)}$ decays
     belonging to the II-a case have branching ratio ${\sim}$
     ${\cal O}(10^{-10})$.
     It is expected to have hundreds or dozens of
     the $B_{c}(2S)$ weak decay events belonging to the I-a, I-b
     and II-a cases with the potential ${\cal O}(10^{11})$ $B_{c}(2S)$ data
     in the coming HL-LHC experiment, which might be observed
     and used to identify the $B_{c}(2S)$ mesons.
     Interestingly, the $B_{c}(2S)$ ${\to}$ $B_{s}\,{\pi}$,
     $B_{s}\,{\rho}$ decays have the top priority in searching
     for the nonleptonic $B_{c}(2S)$ weak decays at the HL-LHC
     experiment, where the characteristic charged ${\pi}^{\pm}$
     and ${\rho}^{\pm}$ particles with a definite momentum
     ${\sim}$ $1.33$ GeV in the rest $B_{c}(2S)$ frame should be
     easily and unambiguously identified by various
     fine-resolution sub-detectors with high
     detection efficiency.

     (3)
     The charmonium particles have often and widely been used to tag
     and reconstruct the weak decay events of the ground $B_{c}(1S)$
     meson at experiments, such as the semileptonic $B_{c}(1S)$
     ${\to}$ $J/{\psi}(1S)\,{\ell}\,{\nu}_{\ell}$ decays and hadronic
     $B_{c}(1S)$ ${\to}$ ${\psi}(nS)\,{\pi}$, ${\psi}(nS)\,{\pi}\,{\pi}$,
     ${\psi}(nS)\,{\pi}\,{\pi}\,{\pi}$,
     ${\psi}(nS)\,K\,{\pi}\,{\pi}$, ${\psi}(nS)\,K\,K\,{\pi}$,
     $J/{\psi}\,D$, ${\chi}_{cJ}\, {\pi}$, ....
     decays ($n$ $=$ $1$ and $2$) \cite{PhysRevD.110.030001}.
     However, the numbers in Table \ref{br:b-decay} show that
     branching ratios for the $B_{c}(2S)$ ${\to}$ ${\eta}_{c}(1S,2S)\,P$,
     ${\psi}(1S,2S)\,P$ decays are significantly small,
     about ${\cal O}(10^{-12})$ for $P$ $=$ ${\pi}$ and
     ${\cal O}(10^{-13})$ for $P$ $=$ $K$.
     These $B_{c}(2S)$ weak decays seem to be exceedingly hard to
     measure, even at the future HL-LHC experiments.
     An observation of the nonleptonic $B_{c}(2S)$ weak decays
     into the final states containing one charmonium meson might
     be a hint of new physics beyond SM.

     (4)
     Due to the relation of decay constants, $f_{\rho}$ $>$ $f_{\pi}$
     and $f_{K^{\ast}}$ $>$ $f_{K}$, there are some hierarchy
     relations among the branching ratios, such as,
     \begin{equation}
    {\cal B}r( B_{c}(2S) {\to} B_{q}^{0} {\rho}) \, > \,
    {\cal B}r( B_{c}(2S) {\to} B_{q}^{0} {\pi})
     \label{br-bq-rho-pi},
     \end{equation}
     \begin{equation}
    {\cal B}r( B_{c}(2S) {\to} B_{q}^{0} K^{\ast}) \, > \,
    {\cal B}r( B_{c}(2S) {\to} B_{q}^{0} K)
     \label{br-bq-Kv-K},
     \end{equation}
     for the same subscript $q$ $=$ $u$, $d$ and $s$.
     Because of the hierarchy of volume of phase space resulting
     from the mass relation, $m_{{\eta}_{c}(1S)}$
     $<$ $m_{{\psi}(1S)}$ $<$ $m_{{\eta}_{c}(2S)}$
     $<$ $m_{{\psi}(2S)}$,
     there is an approximate relationship
     among the branching ratios,
     \begin{eqnarray} & &
    {\cal B}r( B_{c}(2S) {\to} {\eta}_{c}(1S) P) \nonumber \\ & {\ge} &
    {\cal B}r( B_{c}(2S) {\to}     {\psi}(1S) P) \nonumber \\ & {\ge} &
    {\cal B}r( B_{c}(2S) {\to} {\eta}_{c}(2S) P) \nonumber \\ & {\ge} &
    {\cal B}r( B_{c}(2S) {\to}     {\psi}(2S) P)
     \label{br-cc-p},
     \end{eqnarray}
     for the same pseudoscalar $P$ $=$ ${\pi}$ and $K$ meson.

     (5)
     Although it has been proofed that the QCDF coefficients $a_{1,2}$
     in Eq.(\ref{qcdf-a1}) and Eq.(\ref{qcdf-a2}) are the
     renormalization scale independent at the order of ${\alpha}_{s}$
     \cite{PhysRevD.64.014036,PhysRevD.77.074013},
     the main theoretical uncertainties of branching ratios for the
     $B_{c}(2S)$ ${\to}$ $B_{q}\,P$, $B_{q}\,V$ decays come from
     the choice of renormalization scale, which are expected to
     be reduced by considering the contributions of the further
     higher ${\alpha}_{s}$ order in the future.
     For the $B_{c}(2S)$ ${\to}$ ${\eta}_{c}(1S,2S)\,P$,
     ${\psi}(1S,2S)\,P$ decays, the theoretical uncertainties
     of branching ratios come mainly from the CKM factors.
     In addition, we would like to point out that in practice, one
     should not be too serious about the branching ratios in Table
     \ref{br:c-decay} and Table \ref{br:b-decay},
     because at least the total decay width of the $B_{c}(2S)$
     meson has yet not be determined experimentally.
     These numbers in Table \ref{br:c-decay} and Table \ref{br:b-decay}
     are just an order-of-magnitude estimation of the branching ratios.
     In order to reduce the uncertainties, a common way is to
     introduce the ratio of branching ratios.
     Here, we will give some ratios of branching ratios to be
     tested in the future experiments, for example,
     \begin{equation}
     \frac{ {\cal B}r( B_{c}(2S) {\to} B_{s}^{0} K^{\ast} ) }
          { {\cal B}r( B_{c}(2S) {\to} B_{s}^{0} {\rho}   ) }
     \ {\approx} \
     \frac{ {\cal B}r( B_{c}(2S) {\to} B_{d}^{0} K^{\ast} ) }
          { {\cal B}r( B_{c}(2S) {\to} B_{d}^{0} {\rho}   ) }
     \ {\approx} \
    {\vert} V_{us} {\vert}^{2} \, \frac{ f_{K^{\ast}}^{2} }{ f_{\rho}^{2} }
     \ {\approx} \ 0.05
     \label{ratio-rho-KV},
     \end{equation}
     \begin{eqnarray} & &
     \frac{ {\cal B}r( B_{c}(2S) {\to} B_{s}^{0} K     ) }
          { {\cal B}r( B_{c}(2S) {\to} B_{s}^{0} {\pi} ) }
     \ {\approx} \
     \frac{ {\cal B}r( B_{c}(2S) {\to} B_{d}^{0} K     ) }
          { {\cal B}r( B_{c}(2S) {\to} B_{d}^{0} {\pi} ) }
     \nonumber \\ & {\approx} &
     \frac{ {\cal B}r( B_{c}(2S) {\to} {\eta}_{c}(1S) K     ) }
          { {\cal B}r( B_{c}(2S) {\to} {\eta}_{c}(1S) {\pi} ) }
     \ {\approx} \
     \frac{ {\cal B}r( B_{c}(2S) {\to} {\eta}_{c}(2S) K     ) }
          { {\cal B}r( B_{c}(2S) {\to} {\eta}_{c}(2S) {\pi} ) }
     \nonumber \\ & {\approx} &
     \frac{ {\cal B}r( B_{c}(2S) {\to} {\psi}(1S) K     ) }
          { {\cal B}r( B_{c}(2S) {\to} {\psi}(1S) {\pi} ) }
     \ {\approx} \
     \frac{ {\cal B}r( B_{c}(2S) {\to} {\psi}(2S) K     ) }
          { {\cal B}r( B_{c}(2S) {\to} {\psi}(2S) {\pi} ) }
     \nonumber \\ & {\approx} &
    {\vert} V_{us} {\vert}^{2} \, \frac{ f_{K}^{2} }{ f_{\pi}^{2} }
     \ {\approx} \ 0.07
     \label{ratio-pi-K}.
     \end{eqnarray}

     \section{Summary}
     \label{sec04}
     The excited $B_{c}(2S)$ meson has been discovered and
     confirmed experimentally for a few years.
     The $B_{c}(2S)$ meson decays dominantly via the hadronic
     cascades and electromagnetic transitions.
     The $B_{c}(2S)$ meson decays via the weak interaction
     are also allowable in SM, but with a tiny share.
     The identification of the $B_{c}(2S)$ meson faces
     enormous challenges and many difficulties at experiments.
     It is expected that there is a promising prospect of
     ${\cal O}(10^{11})$ $B_{c}(2S)$ meson available
     at the coming HL-LHC experiments.
     Motivated by the huge statistical $B_{c}(2S)$ data,
     the potential of searching for the $B_{c}(2S)$ meson weak
     decays is investigated with the popular QCDF approach.
     Using the form factors obtained with the Wirbel-Stech-Bauer
     approach, the branching ratios for both the $B_{c}(2S)$ ${\to}$
     $B_{q}\,P$, $B_{q}\,V$ decays induced by the $c$ ${\to}$ $q$ quark
     transition and the $B_{c}(2S)$ ${\to}$ ${\eta}_{c}(1S,2S)\,P$,
     ${\psi}(1S,2S)\,P$ decays induced by the $b$ ${\to}$ $c$
     quark transition are estimated theoretically.
     It is found that branching ratios of the hadronic $B_{c}(2S)$
     weak decays into final states containing one bottomed meson
     can reach up to ${\cal O}(10^{-9})$ for the I-a case,
     and ${\cal O}(10^{-10})$ for the II-a and I-b cases,
     and might be measurable.
     The $B_{c}(2S)$ weak decays into final states containing one
     charmonium are severely suppressed by the CKM factors, and have
     significantly small branching ratios, ${\cal O}(10^{-12})$
     and less, which might be outside the future experimental
     detectability.
     It is hoped that our estimation on the hadronic $B_{c}(2S)$
     weak decays can provide a ready reference for the future
     experimental study.

     \section*{Acknowledgments}
     The work is supported by the National Natural Science Foundation
     of China (Grant No. 12275068),
     National Key R\&D Program of China (Grant No. 2023YFA1606000),
     Natural Science Foundation of Henan Province
     (Grant Nos. 252300421491, 242300420250).

     \end{document}